\documentclass[11pt,a4paper]{article}
\usepackage{amsmath,amsthm,amssymb,epsfig,latexsym}
\newcommand{\be}[1]{\begin{equation}\label{#1}}
\newcommand{\ba}[1]{\begin{eqnarray}\label{#1}}
\newcommand{\ee}{\end{equation}}
\newcommand{\ea}{\end{eqnarray}}

\newcommand{\num}{\\\rule{0pt}{20pt}}
\newcommand{\numa}[1]{\\\rule{0pt}{#1pt}}
\newcommand{\dis}{\displaystyle}
\newcommand{\eq}[1]{(\ref{#1})}

\newcommand{\ipt}{\frac{i\pi}3}
\newcommand{\ips}{\frac{i\pi}6}
\begin{document}
\begin{flushright}
LPENSL-TH-05\\
\end{flushright}
\par \vskip .1in \noindent

\vspace{24pt}

\begin{center}
\begin{LARGE}
{\bf Exact results for the $\sigma^z$ two-point function of the $XXZ$ chain at
$\Delta=1/2$}
\end{LARGE}

\vspace{50pt}

\begin{large}

{\bf N.~Kitanine}\footnote[1]{LPTM, UMR 8089 du CNRS, Universit\'e de
Cergy-Pontoise, France, kitanine@ptm.u-cergy.fr},~~
{\bf J.~M.~Maillet}\footnote[2]{ Laboratoire de Physique, UMR 5672 du CNRS, ENS
Lyon,  France,
 maillet@ens-lyon.fr},~~
{\bf N.~A.~Slavnov}\footnote[3]{ Steklov Mathematical Institute, Moscow,
Russia, nslavnov@mi.ras.ru},~~
{\bf V.~Terras}\footnote[4]{LPTA, UMR 5207 du CNRS, Montpellier, France,
terras@lpta.univ-montp2.fr} \par

\end{large}

\vspace{80pt}

\centerline{\bf Abstract} \vspace{1cm}
\parbox{12cm}{\small We propose a new multiple integral representation for the
correlation function $\langle\sigma_1^z\sigma_{m+1}^z\rangle$ of the $XXZ$
spin-$\frac12$ Heisenberg chain in the disordered regime. We show that for
$\Delta=1/2$ the integrals can be separated and computed exactly. As an example
we give the explicit results up to the lattice distance $m=8$. It turns out
that the answer is given as integer numbers divided by $2^{(m+1)^2}$.}
\end{center}

\newpage
In this paper we study the correlation functions of the $XXZ$ spin-$\frac12$
Heisenberg chain \cite{Hei28}. The Hamiltonian of this model is given by
\be{H0}
 H=\sum_{m=1}^{M}\left\{
    \sigma^x_{m}\sigma^x_{m+1}+\sigma^y_{m}\sigma^y_{m+1}
    +\Delta(\sigma^z_{m}\sigma^z_{m+1}-1)\right\},
\ee
where $\Delta$ is the anisotropy parameter and $\sigma^{x,y,z}_{m}$ are local
spin operators associated with the $m$-th site of the chain. The boundary
conditions are periodical.

In 2001 Razumov and Stroganov  \cite{RazS01} observed that, at the special value of
the anisotropy parameter $\Delta=1/2$, the model exhibits a series of remarkable
properties. It was conjectured in particular that, in the thermodynamic limit,
the probability to find a ferromagnetic string of length $m$ in the
antiferromagnetic ground state is proportional to the number of  alternating
sign matrices of size $m\times m$. This conjecture was proved in
\cite{KitMST02c} using multiple integral representations of the correlation
functions \cite{JimMMN92}, \cite{JimM96}, \cite{KitMT00}, \cite{KitMST02a}.
We use here the same approach to study the two-point correlation function
of the third components of local spin.

Following the strategy of \cite{IzeK85}, \cite{KitMST02a}, we introduce the
operator
\be{GFdefQ}
  Q_\kappa(m)
  =\prod_{n=1}^m\left(\frac{1+\kappa}2
  +\frac{1-\kappa}2\cdot\sigma_n^z\right),
\ee
where $\kappa$ is a complex number. The ground state expectation value of this
operator is the generating function for the correlation functions of the third
components of spin, namely,
\be{2-point}
 \langle\sigma_1^z\sigma_{m+1}^z\rangle=2\left.\frac{\partial^2}
 {\partial\kappa^2}D^2_m \langle Q_\kappa(m)\rangle\right|_{\kappa=1}-1,
\ee
where $D^2_m$ means the second lattice derivative.

To study the generating function $\langle Q_\kappa(m)\rangle$ in the
thermodynamic limit, we use the following multiple integral representation:
\begin{multline}\label{new-ser}
 \langle Q_\kappa(m)\rangle
 =\sum_{n=0}^m\frac{\kappa^{m-n}}{n!(m-n)!}
 \oint\limits_{\Gamma\{-i\zeta/2\}} \frac{d^{m}z}{(2\pi i)^{m}}
  \int\limits_{\mathbb{R}-i\zeta}d^{n}\lambda
 \int\limits_{\mathbb{R}}d^{m-n}\lambda\cdot\prod_{j=1}^m
 \varphi^m(z_j)\varphi^{-m}(\lambda_j)
  \\
 \times
  \prod_{j=1}^{n}\left\{t(z_j,\lambda_j)
  \prod_{k=1}^{m}\frac{\sinh(z_j-\lambda_k-i\zeta)}{\sinh(z_j-z_k-i\zeta)}
  \right\}
 \prod_{j=n+1}^{m}\left\{t(\lambda_j,z_j)
 \prod_{k=1}^{m}\frac{\sinh(\lambda_k-z_j-i\zeta)}{\sinh(z_k-z_j-i\zeta)}
  \right\}\\
 \times\prod_{j=1}^{m}\prod_{k=1}^{m}
  \frac{\sinh(\lambda_k-z_j-i\zeta)}{\sinh(\lambda_k-\lambda_j-i\zeta)}
 \cdot{\det}_{m}\left(\frac{i}{2\zeta\sinh\frac\pi\zeta(\lambda-z)}\right).
 \end{multline}
Here,
\be{phi}
 \Delta=\cos\zeta,\qquad
 t(z,\lambda)=\frac{-i\sin\zeta}{\sinh(z-\lambda)\sinh(z-\lambda-i\zeta)},\qquad
 \varphi(z)=\frac{\sinh(z-i\frac\zeta2)}{\sinh(z+i\frac\zeta2)},
\ee
and the integrals over the variables $z_j$ are taken with respect to a closed contour
$\Gamma$ which surrounds the point $-i\zeta/2$ and does not contain any other
singularities of the integrand.

The equation \eq{new-ser} can be derived from the one obtained for $\langle
Q_\kappa(m)\rangle$ in \cite{KitMST02a} by expanding the integrand into power
series over $\kappa$ with forthcoming  moving  a part of the integration
contours for variables $\lambda_j$ to the lower half-plane.

%%%%%%%%%%%%%%%%%%%%%%%%%%%%%%%%%%%%%%%%%%%%%%%%%%%%%%%%%%%%%%%%%%%%%%%%%%%%%%%%
%%%%%%%%%%%%%%%%%%%%%%%%%%%%%%%%%%%%%%%%%%%%%%%%%%%%%%%%%%%%%%%%%%%%%%%%%%%%%%%%
%%%%%%%%%%%%%%%%%%%%%%%%%%%%%%%%%%%%%%%%%%%%%%%%%%%%%%%%%%%%%%%%%%%%%%%%%%%%%%%%
%%%%%%%%%%%%%%%%%%%%%%%%%%%%%%%%%%%%%%%%%%%%%%%%%%%%%%%%%%%%%%%%%%%%%%%%%%%%%%%%
%%%%%%%%%%%%%%%%%%%%%%%%%%%%%%%%%%%%%%%%%%%%%%%%%%%%%%%%%%%%%%%%%%%%%%%%%%%%%%%%

The equation \eq{new-ser} is valid for the homogeneous $XXZ$ chain with arbitrary
$-1<\Delta<1$. If we consider the inhomogeneous $XXZ$ model with
inhomogeneities $\xi_1,\dots,\xi_m$, then one should replace in the
representation \eq{new-ser} the function $\varphi^m$ in the following way:
\be{repl}
 \varphi^m(z)\to\prod_{b=1}^m\frac{\sinh(z-\xi_b-i\zeta)}{\sinh(z-\xi_b)},\qquad
 \varphi^{-m}(\lambda)\to\prod_{b=1}^m\frac{\sinh(\lambda-\xi_b)}
 {\sinh(\lambda-\xi_b-i\zeta)}.
\ee
In order to come back to the homogeneous case, one should set $\xi_k=-i\zeta/2$,
$k=1,\dots,m$ in \eq{repl}.

In the inhomogeneous model, the integration contour $\Gamma$ surrounds the
points $\xi_1,\dots,\xi_m$, and the integrals over $z_j$ are therefore equal to the
sum of the residues of the integrand in these simple poles. Computing these integrals
and setting $\zeta=\pi/3$, we obtain an alternating sum over the permutations of the
set $\{\xi_1,\dots,\xi_m\}$:
\begin{multline}\label{new-ser-12}
 \langle Q_\kappa(m)\rangle
=\frac{(-1)^{\frac{m^2-m}2}}{m!}
 \left(\frac{\sqrt{3}}{2\pi}\right)^m2^{-m^2}
 \prod_{a>b}^m\frac{\sinh3(\xi_a-\xi_b)}{\sinh^2(\xi_a-\xi_b)}
 \\
 \times\sum_{n=0}^m\kappa^{m-n}C_m^n \sum_{p}(-1)^{[p]}
 \prod_{a=1}^m\prod_{b=n+1}^m
 \frac{\sinh(\xi_{p(b)}-\xi_a-\ipt)} {\sinh(\xi_{p(b)}-\xi_a+\ipt)}
\\
 \int\limits_{\mathbb{R}-i\zeta}d^{n}\lambda
 \int\limits_{\mathbb{R}}d^{m-n}\lambda\cdot
  \frac{\prod\limits_{a>b}^m\sinh(\lambda_a-\lambda_b)
 \prod\limits_{j=1}^{n}t(\xi_{p(j)},\lambda_j)
 \prod\limits_{j=n+1}^{m}t(\lambda_j,\xi_{p(j)})}
 {\prod\limits_{a=1}^m\left\{\prod\limits_{b=1}^n
 \sinh(\lambda_a-\xi_{p(b)}-\ipt)
 \prod\limits_{b=n+1}^m\sinh(\lambda_a-\xi_{p(b)}+\ipt) \right\}}.
 \end{multline}
It is easy to see that the double products containing the parameters
$\{\lambda\}$ can be written as a Cauchy determinant with
$\sinh^{-1}(\lambda_j-\xi_{p(k)}-\ipt)$ in the first $n$ columns and
$\sinh^{-1}(\lambda_j-\xi_{p(k)}+\ipt)$ in the remaining ones . Thus, the
integrand can be presented as a determinant of a block-matrix:
\begin{multline}\label{factor}
 \langle Q_\kappa(m)\rangle
 =\frac{1}{2^{m^2}m!}
 \left(\frac{\sqrt{3}}{2\pi}\right)^m
 \prod_{a>b}^m\frac{\sinh3(\xi_a-\xi_b)}{\sinh^3(\xi_a-\xi_b)}
 \\
 \times\sum_{n=0}^m\kappa^{m-n}C_m^n \sum_{p}
 \prod_{a=1}^n\prod_{b=n+1}^m
 \frac{\sinh(\xi_{p(b)}-\xi_{p(a)}-\ipt)\sinh(\xi_{p(a)}-\xi_{p(b)})}
  {\sinh^2(\xi_{p(b)}-\xi_{p(a)}+\ipt)}
\numa{40}
 \int\limits_{\mathbb{R}-i\zeta}d^{n}\lambda
 \int\limits_{\mathbb{R}}d^{m-n}\lambda\cdot
 \det_m\left(\begin{array}{c|c}
 \frac{t(\xi_{p(j)},\lambda_j)}{\sinh(\lambda_j-\xi_{p(k)}-\ipt)}
 & \frac{t(\xi_{p(j)},\lambda_j)}{\sinh(\lambda_j-\xi_{p(k)}+\ipt)}\\
 {}&{}\\
 \hline&{}\\
 \frac{t(\lambda_j,\xi_{p(j)})}{\sinh(\lambda_j-\xi_{p(k)}-\ipt)}
 & \frac{t(\lambda_j,\xi_{p(j)})}{\sinh(\lambda_j-\xi_{p(k)}+\ipt)}
 \end{array}\right).
 \end{multline}
Here the sizes of the blocks in the determinant are respectively: $n\times n$;
$n\times (m-n)$; $(m-n)\times n$; and $(m-n)\times (m-n)$.

We see that the original $m$-fold integral over $\lambda_j$ is now factorized.
Indeed, $\lambda_j$ enters only the $j$-th line of the determinant and, hence,
one can integrate each line separately. Thus, we arrive at
\begin{multline}\label{result-inh}
 \langle Q_\kappa(m)\rangle
 =\frac{3^{m}}{2^{m^2}}
 \prod_{a>b}^m\frac{\sinh3(\xi_{a}-\xi_b)}{\sinh^3(\xi_{a}-\xi_b)}
 \sum_{n=0}^m\kappa^{m-n}\sum_{\{\xi\}=\{\xi_{\gamma_+}\}%
 \cup\{\xi_{\gamma_-}\}\atop{|\gamma_+|=n}}
 \det_m\hat\Phi^{(n)}(\{\xi_{\gamma_+}\},\{\xi_{\gamma_-}\})
 \numa{40}
 \times \prod_{a\in\gamma_+}\prod_{b\in\gamma_-}
 \frac{\sinh(\xi_{b}-\xi_a-\ipt)\sinh(\xi_{a}-\xi_b)}{\sinh^2(\xi_{b}-\xi_a+\ipt)},
 \end{multline}
 with
 \be{h-Phi}
 \hat\Phi^{(n)}(\{\xi_{\gamma_+}\},\{\xi_{\gamma_-}\})=\left(\begin{array}{c|c}
 \Phi(\xi_{j}-\xi_{k})&\Phi(\xi_{j}-\xi_{k}-\ipt)\\
  {}&{}\\
 \hline&{}\\
 \Phi(\xi_{j}-\xi_{k}+\ipt)&\Phi(\xi_{j}-\xi_{k})
 \end{array}\right), \qquad \Phi(x)=\frac{\sinh\frac{x}2}{\sinh\frac{3x}2}.
 \ee
Here the sum is taken with respect to all partitions of the set $\{\xi\}$ into
two disjoint subsets $\{\xi_{\gamma_+}\}\cup\{\xi_{\gamma_-}\}$ of
cardinality $n$ and $m-n$ respectively. The first $n$ lines and columns of the
matrix $\hat\Phi^{(n)}$ are associated with the parameters
$\xi\in\{\xi_{\gamma_+}\}$. The remaining lines and columns are associated with
$\xi\in\{\xi_{\gamma_-}\}$.

Thus, we have obtained an explicit answer for the generating function $\langle
Q_\kappa(m)\rangle$ of the inhomogeneous $XXZ$ model. However, the homogeneous
limit of \eq{result-inh} is not obvious, since each term of the sum in
\eq{result-inh} becomes singular in the limit\footnote[1]{%
Since \eq{result-inh} depends only on the difference $\xi_j-\xi_k$, one can
obtain the homogeneous $XXZ$ chain in the limit $\xi_k\to0$ as well as in the
limit $\xi_k\to-i\zeta/2$.}
 $\xi_k\to0$, $k=1,\dots,m$. In order to check that the sum over partitions in
 the r.h.s. of \eq{result-inh} remains indeed finite in the limit $\xi_k\to0$, one can
 introduce again a set of auxiliary contour integrals:
\begin{multline}\label{result-h}
 \langle Q_\kappa(m)\rangle
 =\frac{(-1)^{\frac{m^2-m}2}3^{m}}{2^{m^2}m!}
 \prod_{a>b}^m\frac{\sinh3(\xi_{a}-\xi_b)}{\sinh(\xi_{a}-\xi_b)}
 \sum_{n=0}^m\kappa^{m-n}C_m^n
 \oint\limits_{\Gamma\{\xi-\ips\}}\frac{d^{n}z}{(2\pi i)^n}
 \oint\limits_{\Gamma\{\xi+\ips\}}\frac{d^{m-n}z}{(2\pi i)^{m-n}}
 \\
 \times\prod_{b=1}^m\left\{\prod_{j=1}^n\frac1{\sinh(z_j-\xi_b+\ips)}
 \prod_{j=n+1}^m\frac1{\sinh(z_j-\xi_b-\ips)}\right\}
 \numa{40}
 \times
 \prod_{a=1}^n\prod_{b=n+1}^m
 \frac{\sinh(z_a-z_b-\ipt)\sinh(z_a-z_b+\ipt)}{\sinh^2(z_a-z_b)}
 \cdot\det_m\Phi(z_j-z_k).
 \end{multline}
Here the integration contours $\Gamma\{\xi\mp\ips\}$ surround the points
$\{\xi-\ips\}$ for $z_1,\dots,z_n$ and $\{\xi+\ips\}$ for $z_{n+1},\dots,z_m$
respectively. It is easy to check that, taking the residues of the integrand in
the simple poles $\{\xi\pm\ips\}$, we immediately reproduce \eq{result-inh}. On
the other hand, one should
simply set $\xi_k=0$, $k=1,\dots,m$ to proceed to the homogeneous limit
in \eq{result-h}. As a result we obtain  poles of order $m$
in the r.h.s. of \eq{result-h}.

Certainly, the remaining integral is of Cauchy type and, after the change of
variables $x_j=e^ {2z_j}$, it reduces to the derivatives of order $m-1$  with
respect to each $x_j$ at $x_1=\cdots=x_n=e^ {\ipt}$ and $x_{n+1}=\cdots=x_m=
e^{-\ipt}$. If the lattice distance $m$ is not too large, the representations
\eq{result-inh}, \eq{result-h} can be successfully used to compute
$\langle Q_\kappa(m)\rangle$ explicitely. As an example we give below the
list of results for $P_m(\kappa)=2^{m^2}\langle Q_\kappa(m) \rangle$ up to
$m=9$:
\be{polynoms}
 \begin{array}{l}
 {\dis P_1(\kappa)=1+\kappa,}\num
 {\dis P_2(\kappa)=2+12\kappa+2\kappa^2,}\num
 {\dis P_3(\kappa)=7+249\kappa+249\kappa^2+7\kappa^3,}\num
 {\dis P_4(\kappa)=42+10004\kappa+45444\kappa^2+10004\kappa^3+42\kappa^4,}\num
 {\dis P_5(\kappa)=429+738174\kappa+16038613\kappa^2+16038613\kappa^3
 +738174\kappa^4+429\kappa^5,}\num
 {\dis P_6(\kappa)=7436+96289380\kappa+11424474588\kappa^2+45677933928\kappa^3
 +11424474588\kappa^4}\num
 {\dis +96289380\kappa^5+7436\kappa^6,}\num
 {\dis P_7(\kappa)= 218348+21798199390\kappa+15663567546585\kappa^2+265789610746333\kappa^3}\num
 {\dis +265789610746333\kappa^4+15663567546585\kappa^5+21798199390\kappa^6+218348\kappa^7,}\num
 {\dis P_8(\kappa)= 10850216+8485108350684\kappa+39461894378292782\kappa^2}\num
 {\dis +3224112384882251896\kappa^3
 +11919578544950060460\kappa^4+3224112384882251896\kappa^5}\num
 {\dis +39461894378292782\kappa^6+ 8485108350684\kappa^7+10850216\kappa^8}\num
 {\dis P_9(\kappa)=911835460+5649499685353257\kappa
 +177662495637443158524\kappa^2}\num
 {\dis+77990624578576910368767\kappa^3+1130757526890914223990168\kappa^4}\num
 {\dis+1130757526890914223990168\kappa^5+77990624578576910368767\kappa^6}\num
 {\dis+177662495637443158524\kappa^7+5649499685353257\kappa^8+911835460\kappa^9.}
 \end{array}
\ee

Using the results \eq{polynoms}, we can calculate the correlation function
$\langle\sigma_1^z\sigma_{m+1}^z\rangle$ up to $m=8$:
 \be{corr-f}
 \begin{array}{l}
 {\dis\langle\sigma_1^z\sigma_2^z\rangle=-2^{-1},}\num
 {\dis\langle\sigma_1^z\sigma_3^z\rangle=7\cdot2^{-6},}\num
 {\dis\langle\sigma_1^z\sigma_4^z\rangle=-401\cdot2^{-12},}\num
 {\dis\langle\sigma_1^z\sigma_5^z\rangle=184453\cdot2^{-22},}\num
 {\dis\langle\sigma_1^z\sigma_6^z\rangle=-95214949\cdot2^{-31},}\num
 {\dis\langle\sigma_1^z\sigma_7^z\rangle=1758750082939\cdot2^{-46},}\num
 {\dis\langle\sigma_1^z\sigma_8^z\rangle=-30283610739677093\cdot2^{-60},}\num
 {\dis\langle\sigma_1^z\sigma_9^z\rangle=5020218849740515343761\cdot2^{-78}.}
 \end{array}
\ee
For $m=2,3$ we reproduce the results \cite{KatSTS03}, \cite{KatSTS04}  where
the two-point correlation functions of the $XXZ$ chain were calculated up to
the lattice distance $m=3$ for general $\Delta$.

One can also  compare the exact results \eq{corr-f} with the values given by
the asymptotic formula suggested in \cite{Luk99}.
\vspace{3mm}
\begin{center}
\begin{tabular}{|c|c|c|}
\hline
m&$\langle\sigma_1^z\sigma_{m+1}^z\rangle\quad\mbox{Exact}$&%
$\langle\sigma_1^z\sigma_{m+1}^z\rangle\quad\mbox{Asymptotics}$\\
\hline
 1&-0.5000000000&-0.5805187860\\
 2&0.1093750000&0.1135152692\\
 3&-0.0979003906&-0.0993588501\\
 4&0.0439770222&0.0440682654\\
 5&-0.0443379157&-0.0444087865\\
 6&0.0249933420&0.0249365346\\
 7&-0.0262668452&-0.0262404925\\
 8&0.0166105110&0.0165641239\\
 \hline
\end{tabular}
\end{center}
\vspace{3mm}
 One can see that
the correlation function $\langle\sigma_1^z\sigma_{m+1}^z\rangle$ approaches
its asymptotic regime very quickly and that, already for $m=4$, the relative precision
is better than $1\%$.

%%%%%%%%%%%%%%%%%%%%%%%%%%%%%%%%%%%%%%%%%%%%%%%%%%%%%%%%%%%%%%%%%%%%%%%%%%%%%%%
%%%%%%%%%%%%%%%%%%%%%%%%%%%%%%%%%%%%%%%%%%%%%%%%%%%%%%%%%%%%%%%%%%%%%%%%%%%%%%%
In conclusion, we would like to comment on the results obtained in this paper.
We have shown that, for the special value of the anisotropy parameter
$\Delta=1/2$, the multiple integral giving the generating function $\langle
Q_\kappa(m)\rangle$ can be factorized and computed explicitly in terms of the
inhomogeneities. It was shown recently in the papers \cite{BooJMST04a},
\cite{BooJMST04b} that a factorization of the multiple integral representations
for the correlation functions of the inhomogeneous $XXZ$ chain is possible for
generic value of $\Delta$. However,  as in  \cite{BooJMST04b}, we see that
it is not straightforward to take the explicit homogeneous limit
of \eq{result-inh} for general $m$, although it is possible to do it for small
distances. One way to obtain explicitly
this homogeneous limit is to come back to the multiple integral of Cauchy type
\eq{result-h}, but with non-factorized integrand.

Concerning the exact numerical formulas we obtained for the generating
function $\langle Q_\kappa(m)\rangle$ up to $m=9$, we would like to mention
the remarkable fact
that all the coefficients of the polynomials $P_m(\kappa)$ turned out to be
integers. We conjecture that this property holds for all $m$. In particular,
the highest and lowest coefficients of $P_m(\kappa)$ correspond to the
emptiness formation probability and, hence, are equal to the number of
alternating sign matrices of the size $m\times m$ \cite{RazS01},
\cite{KitMST02c}. For the other coefficients, this property should be related to the
structure of the ground state at $\Delta=1/2$ conjectured by Razumov and
Stroganov in \cite{RazS01} (see also \cite{RazS04}, \cite{GieBNM02},
\cite{GieBNM04}). It would be very interesting to reveal the combinatorial
nature of these coefficients.

%
%%%%%%%%%%%%%%%%%%%%%%%%%%%%%%%%%%%%%%%%%%%%%%%%%%%%%%%%%%%%%%%%%%%%%%%%%%%%%%%%%%%%%%
\section*{Acknowledgements}
J. M. M., N. S. and V. T. are supported by CNRS. N. K., J. M. M., V. T. are
supported by the European network EUCLID-HPRNC-CT-2002-00325.  J. M. M. and
N.S. are supported by INTAS-03-51-3350. N.S. is supported by the French-Russian
Exchange Program, the Program of RAS Mathematical Methods of the Nonlinear
Dynamics, RFBR-05-01-00498, Scientific Schools 2052.2003.1. N. K, N. S. and V.
T. would like to thank the Theoretical Physics group of the Laboratory of
Physics at ENS Lyon for hospitality, which makes this collaboration possible.

%%%%%%%%%%%%%%%%%%%%%%%%%%%%%%%%%%%%%%%%%%%%%%%%%%%%%%%%%%%%%%%%%%%%%%%%%%%%%%

\end{document}